\documentstyle[twocolumn,prl,aps,epsfig,floats]{revtex}
\begin{document}
\draft

\twocolumn[\hsize\textwidth\columnwidth\hsize\csname @twocolumnfalse\endcsname

\newcommand{\ibf}{\mbox{\boldmath $f$}}
\title{
Crystal Field Triplets: A New Route to Non-Fermi Liquid Physics
}

\author{
Mikito Koga$^1$, Gergely Zar{\'a}nd$^{1,2}$, and Daniel L. Cox$^1$
}

\address{
$^1$Department of Physics, University of California Davis, CA 95616\\ 
$^2$Research Group of the Hungarian Academy of Sciences, Institute of Physics,
TU Budapest, H-1521
}
\date{\today}
\maketitle

\begin{abstract}
A model for crystal field triplet ground states on rare earth or actinide
ions with dipolar and quadrupolar couplings to conduction electrons is
studied for the first time with renormalization group methods.
The quadrupolar coupling leads to a new nontrivial, non-Fermi liquid fixed
point, which survives in an intermediate valence Anderson model.
The calculated magnetic susceptibility displays one parameter scaling,
going as $T^{-\alpha}$ ($\alpha \approx 0.4$) at intermediate temperatures,
reminiscent of the non-Fermi liquid alloy UCu$_{5-x}$Pd$_x$.
\end{abstract}

\pacs{PACS numbers: 75.20.Hr, 71.10.Hf, 71.27.+a, 72.15.Qm}

]
\narrowtext
Recent data for many ``heavy fermion'' Ce or U based compounds and alloys display
diverging low-temperature magnetic susceptibility $\chi(T)\sim T^{-\alpha}$ and
electronic specific heat coefficients $C_{el}/T=\gamma(T) \sim T^{-\alpha}$
or $\sim -\ln(T)$, unlike 
those of a Fermi liquid ($\chi(0),\gamma(0) ~ constant$) \cite{Sakai99}.
A number of theoretical
scenarios have arisen to explain this non-Fermi lquid (NFL) physics, which
broadly fall into two categories: (1) {\it Localized}, including 
models associated with peculiar symmetry allowed
interactions between $f$-electron moments and conduction electrons
\cite{Nozi80,Cox87,Koga95,Shimi98,Cox98} or
a disorder induced distribution of Kondo scales
\cite{Mira97,Bernal95}.
Among the first class of models, multi-channel Kondo Hamiltonians employing
couplings between localized magnetic or orbital doublets (induced
by crystalline electric field [CEF] splittings) and conduction
electrons have been extensively studied\cite{Cox87,Koga95,Shimi98,Cox98}.
(2) {\it Extended}, in which the NFL behavior is driven by coupling of
electrons to low-lying modes induced by intersite $f$-moment
interactions in proximity to a quantum critical point
\cite{Millis93,Castro98}.
It is of considerable interest
to sort out the applicability of these differing scenarios to real
materials.

Of particular interest is the alloy system UCu$_{5-x}$Pd$_x$ \cite{Andra98}
which displays NFL behavior in the range $1\le x \le 2.5$ that
has been described in terms of localized disordered Kondo physics
\cite{Mira97,Bernal95}
or Griffiths' phase theory \cite{Castro98}.  This alloy appears to
display two separate NFL regimes as a function of temperature: (1) An
``impurity'' regime independent of $x$
for $80K \le T \le 300K$ with $\chi(T)\sim T^{-1/3}$
and $\chi''(\vec Q,\omega,T) \sim \omega^{-1/3}$ for $\hbar\omega \ge k_BT$ (there
appears to be negligible $Q$ dependence to $\chi''$) \cite{Aron95,Maple}.  (2)
A low temperature ``lattice'' regime in which $\chi(T) \sim \gamma(T) \sim T^{-\alpha(x)}$
and for which some evidence exists for intersite interaction effects
\cite{Maple}.
The ``impurity'' regime data is not compatible with any plausible
multi-channel Kondo model assuming ground state magnetic or orbital
doublet levels on the U ions.  The picture is further complicated by
analysis of photoemission data which suggests the U ions are in the
mixed valent regime, possessing a nearly 50-50 mix of ground state
weight in the $f^3$ and $f^2$ configurations \cite{Allen}. This is 
problematic in that the Kondo effect has
been conventionally studied only for nearly integral valent ions. 
\par

In this paper, we present a Kondo model for uranium ions with a 
CEF triplet ground state that is allowed in cubic symmetry.  This model
features an effective spin 1 local moment coupled via magnetic dipole
and electric quadrupole interactions to one band of 
effective spin 3/2 conduction electrons.  
The model displays a new stable NFL fixed point (FP) at 
low temperatures which
is robust even in the mixed valent regime of the more fundamental
Anderson model.
We find three different power-law regions for the magnetic susceptibility
$\chi(T)$:
a characteristic $T^{-\alpha}$ ($\alpha \approx 0.4$) dependence
in the intermediate temperature region over two decades, a Curie-Weiss
law $T^{-1}$ at higher temperatures, and a universal power-law behavior
$T^{-2/3}$ at lower temperatures.
Despite the differing temperature regimes, a surprising
one parameter scaling emerges for $\chi(T)$.
The quadrupolar coupling is demonstrated to be relevant using
multiplicative and numerical renormalization group (NRG) methods
\cite{Cox98,Wilson75}.
We compare the FP properties with 
those of the unstable  FP reached in the
absence of the quadrupolar coupling, and examine the FP structure
in the presence of uniaxial symmetry breaking fields. We argue that the
intermediate and low temperature regimes may be relevant to the physics
of UCu$_{5-x}$Pd$_x$.
\par

Assuming a dominantly tetravalent (5$f^2$) U ion for the moment, 
the Hund's rule ground state has total angular momentum $J=4$, which is
split into a quadrupolar ($\Gamma_3$) doublet, two magnetic triplets
($\Gamma_4$,$\Gamma_5$) and a singlet ($\Gamma_1$) under the action of
the cubic CEF.  The possible ground states, which are accessed by
varying the two parameters of the crystal field Hamiltonian, are
$\Gamma_3$, $\Gamma_5$, or $\Gamma_1$ \cite{Lea62}.  As argued elsewhere,
the $\Gamma_3$ ground state will give rise to the two-channel
quadrupolar Kondo effect upon coupling to conduction
electrons\cite{Cox87,Koga95,Cox98}.
The two-channels arise from coupling to a local quartet ($\Gamma_8$) of
conduction electrons which may be described as a tensor product of states
with two magnetic labels and two orbital ($\Gamma_3$) labels.
However, the analysis of the $\Gamma_8$ tensors reveals two possible
``dipole'' operators:
in addition to the set of effective magnetic doublets labeled by the
orbital indices, it can be written as a single $s=3/2$ manifold.
We find that the latter dominates the coupling to the triplet ground state,
and this model has not been studied before.
\par

The $f^2$ triplet has both magnetic and quadrupolar moments, and can be 
represented by a pseudo-spin $S = 1$ which can couple both to the above
mentioned $\Gamma_8$ quartet of conduction electrons and to a local
doublet ($\Gamma_6$ or $\Gamma_7$).  When the coupling to the quartet is
larger, the corresponding Kondo Hamiltonian can be written as
\begin{eqnarray}
H & = & \sum_{km} \varepsilon_{k} c_{km}^{\dag} c_{km}
\mbox{} + \sum_{kk'mm'} c_{k'm'}^{\dag} c_{km}
\nonumber \\
& \times & \left[ J_{\rm D} ({\vec S}_{\rm c})_{m'm} \cdot {\vec S}
\mbox{} + J_{\rm Q} ({\vec Q}_{\rm c})_{m'm} \cdot {\vec Q} \right].
\label{eq:1}
\end{eqnarray}
Here the $S_{\rm c} = 3/2$ ($m,m' = \pm 1/2, \pm 3/2$) spin operator
represents the four-fold $\Gamma_8$ states of the conduction
electrons and a potential scattering term is neglected.
The quadrupolar operators are given by $\{Q^i,\; i= 1,..,5\} = 
\{S_yS_z + S_zS_y,\; S_zS_x + S_xS_z,\; S_xS_y + S_yS_x,\; S_x^2 - S_y^2,\; 
(2S_z^2 - S_x^2 - S_y^2)/\sqrt{3}\;\}$.
The conduction electron with wave number $k$ and pseudo-spin
$m$ has kinetic energy $\varepsilon_k$ and is created (annihilated)
by the operator $c_{km}^{\dag}$ ($c_{km}$).
In the limit of small hybridization between the conduction band and
the $f$-orbitals, this Kondo Hamiltonian (\ref{eq:1}) can be directly derived
from an Anderson Hamiltonian where we restrict ourselves to the valence
fluctuation between the $5f^1$$\Gamma_7$ and $5f^2$$\Gamma_4$ ($\Gamma_5$)
states. In this case, we obtain a coupling ratio $J_{\rm D} / J_{\rm Q} = 2$
and a marginally irrelevant potential scattering with amplitude $J_{\rm D}$.
Our NRG calculations show that even including the $5f^3$ configuration 
and extending the Anderson model parameters to the mixed-valent regime,
the Kondo model (\ref{eq:1}) describes a stable FP \cite{Koga99}.
Therefore, at low enough temperatures,
we can use this model to study the realistic Kondo effect corresponding to
this FP.
\par

The relevance of the quadrupolar coupling $J_{\rm Q}$ in Eq.~(\ref{eq:1}) can
be immediately seen from a multiplicative renormalization group procedure,
valid in the weak coupling regime.
After a straightforward but lengthy calculation we derive the following RG
equations:
\begin{eqnarray}
{dj_{\rm D} / dx} &= & (j_{\rm D}^2+12 j_{\rm Q}^2)(1-5 j_{\rm D})\; \nonumber \\
{dj_{\rm Q} / dx} & = &6j_{\rm D} j_{\rm Q} - 36 j_{\rm Q}^3 - 15 j_{\rm D}^2 j_{\rm Q} \;.
\label{eq:scaling}
\end{eqnarray} 
Here $x = {\rm ln}(E_{\rm F}/\omega)$ denotes the scaling variable (with 
$E_{\rm F}$, the Fermi energy, and $\omega$, the characteristic energy scale), 
and we have introduced the dimensionless couplings
$j_{Q} = \varrho_0 J_{\rm Q}$ and
$j_{\rm D} =\varrho_0 J_{\rm D}$ with $\varrho_0$,
the density of states at the Fermi surface.
The flow diagrams obtained from a numerical solution of
Eq.~(\ref{eq:scaling}) are shown in Fig.~1.
\begin{figure}
\begin{center}
\psfig{figure=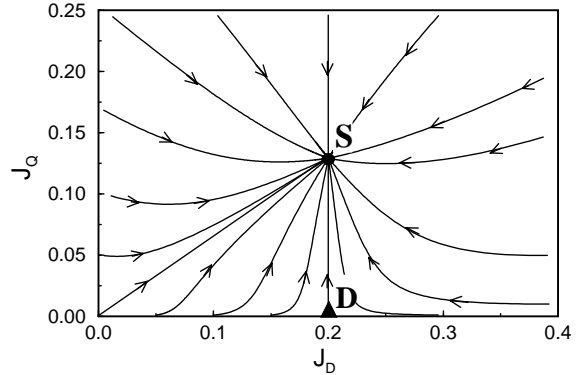,width=7.5cm}
\end{center}
\vspace*{0.05cm}
\caption{
Scaling trajectories obtained from Eq.~(\protect{\ref{eq:scaling}}).
}
\end{figure}
In the absence of quadrupolar exchange the 
model scales to the dipolar FP "D" at ($j_{\rm Q} = 0$, $j_{\rm D} = 1/5$).
This FP has been shown to be characterized by a critical exponent
$\Delta = 1/6$ associated with its spin sector, which can be mapped to the
spin sector of the  10-channel Kondo problem \cite{Kim97,Seng97}.
Obviously, this dipolar FP is unstable to quadrupolar perturbations and for
any non-zero $j_{\rm Q}$ it flows to a new FP "S" at
($j_{\rm D} = 1/5$, $j_{\rm Q} = \sqrt{1/60}$).
\par

At the FP "S" the ratio  $j_{\rm Q}/j_{\rm D}$ takes the value  $j_{\rm Q}/j_{\rm D} = \sqrt{5/12}$, 
and the interaction part of  the Hamiltonian can be written in the following 
pseudo-$SU(3)$ invariant form:
\begin{equation}
H_{\rm int} = J \sum_{kk'mm' } \sum_{i = 1} ^8 \lambda^i\; c^\dagger_{km}
({\lambda}^i_{\rm c})_{mm'} c_{k'm'}\;,
\label{eq:FPstructure}
\end{equation}
where the $\lambda^i$'s denote the 3$\times$3 Gell-Mann matrices
satisfying the $SU(3)$ Lie algebra  $[\lambda^i, \lambda^j] = 
2 i f^{ijk}\lambda^k$, and can be easily expressed in terms of the 
spin one impurity operators.
The 4$\times$4 matrices  $\lambda^i_{\rm c}$ are constructed 
from the conduction electron spin  operators $S_{\rm c}$, and 
satisfy a "pseudo-$SU(3)$" Lie algebra:  $[\lambda^i_{\rm c}, 
\lambda^j_{\rm c}] =  2 i f^{ijk}\lambda^k_{\rm c} + \mbox{octupolar 
~terms}$.  While these latter terms, which arise from the commutators,
spoil the $SU(3)$ symmetry of the local triplet, they cannot couple to
the impurity, and are hence irrelevant in a renormalization group and
general sense.  Since  both the leading 
(second order) and next leading logarithmic 
(third order) scaling equations result in the FP 
structure of Eq.~(\ref{eq:FPstructure}), we believe that this
result is universal and independent of the weak coupling approximation.
\par

Unfortunately, the pseudo-symmetry found is not strong enough for
the usual characterization of the FP by boundary conformal 
field theory (CFT) \cite{Affl90};
in particular, it is impossible to absorb the
impurity spin in the conduction electron currents without 
violating the Kac-moody algebra of the conduction electrons. 
However, we can study the properties of the novel FP by using the NRG.
Following Wilson \cite{Wilson75}, we rewrite Eq.~(\ref{eq:1}) as
\begin{eqnarray}
H_{N+1} & = & \Lambda^{1/2} H_N
\mbox{} + \sum_m (f_{N+1,m}^{\dag} f_{Nm} + {\rm H.c.}),
\nonumber \\
H_0 & = & \sum_{mm'} \sum_{{\rm T}={\rm S},\rm{Q}} f_{0,m'}^{\dag} f_{0,m}
\left[ \tilde{J}_{\rm T} ({\vec T}_{\rm c})_{m'm} \cdot {\vec T} \right],
\end{eqnarray}
where $H_0$ represents the effective exchange interaction on the impurity site,
$f_{Nm}^{\dag}$ ($f_{Nm}$) creates (annihilates) a conduction electron in the
logarithmic discretized band, and
$\tilde{J} = 2J \varrho_0 \Lambda^{-1/2} / (1 + \Lambda^{-1})$ with
$\Lambda$, the discretization parameter.  We follow the usual procedure
and iteratively diagonalize $H_{N+1}$ to probe the system size on a
scale of order $k_{\rm F}^{-1} \Lambda^{N/2}$ and temperature of order
$T_{\rm F}\Lambda^{-N/2}$.  
\par

\begin{figure}
\begin{center}
\psfig{figure=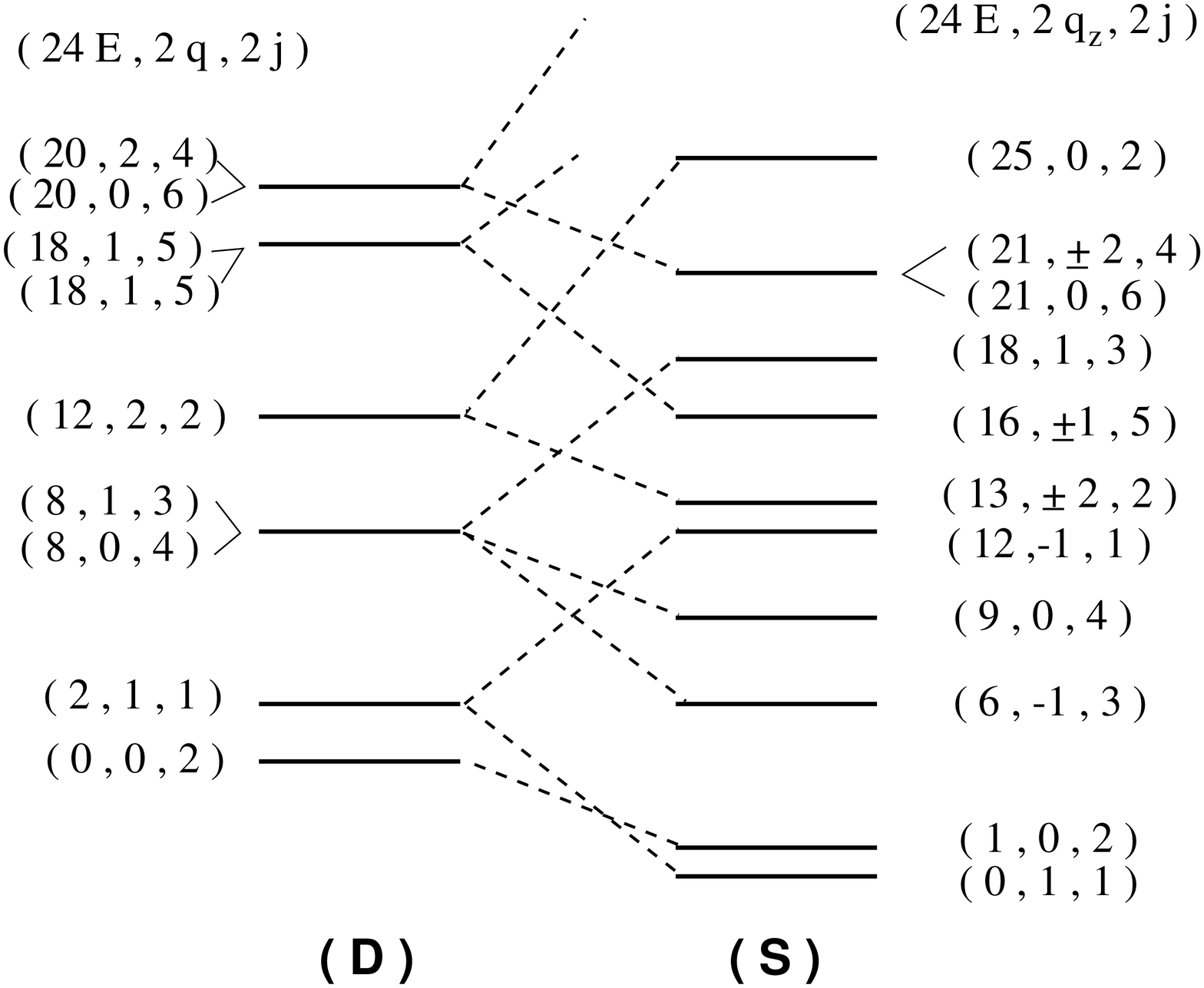,width=7.5cm}
\end{center}
\vspace*{0.05cm}
\caption{
Finite-size energy spectrum of the fixed points D and S of Fig.~1.
The energy $E$ is measured in units of  $\pi / L$ ($v_{\rm F} = 1$) with $L$,
the system size.
(D) For $J_{\rm Q} = 0$, each energy level is labeled by axial charge $q$ and
total spin $j$.
(S) For $J_{\rm Q} \neq 0$, $q$ is not a good quantum number since
the charge $SU(2)$ symmetry is broken by the quadrupolar interaction.
}
\end{figure}
Fig.~2 shows the finite-size energy spectrum obtained at FP's "D" and "S".
The  $J_{\rm Q} = 0$ spectrum in the sector with dipolar coupling only 
("D") coincides with the exact 
CFT spectrum for the model of the impurity spin coupled to spin 1/2
electrons \cite{Kim97}.
In the sector with a finite value of $J_{\rm Q}$ ("S"),
we show the energy spectrum we anticipate for a CFT of the model inferred
from our NRG calculations (even iteration) for initial small values of
$\tilde{J}_{\rm D}$ and $\tilde{J}_{\rm Q}$.
The axial charge operator  ${\vec q}$  in Fig.~2 is defined by
$q_+ = \sum_{n=0}^{\infty} (-1)^n
(f_{n,3/2}^{\dag} f_{n,-3/2}^{\dag} - f_{n,1/2}^{\dag} f_{n,-1/2}^{\dag})$, and
$q_z = {1 \over 2} \sum_{n=0}^{\infty} \sum_m
\left(f_{nm}^{\dag} f_{nm} - {1 \over 2} \right)$,
and  satisfies the $SU(2)$ Lie algebra, 
$[q_z,q_{\pm}] = \pm q_{\pm}$ and $[q_+,q_-] = 2q_z$ \cite{Kim97}.
It is easy to show that the relevant quadrupolar coupling breaks the
axial charge $SU(2)$ 
symmetry down to $U(1)$, while it conserves that of the
total spin $j$.
This symmetry breaking, however, cannot be described by a simple 
phase shift, as is often the case: as 
shown in Fig.~2, some originally degenerate 
$SU(2)$ axial charge multiplets are split at the new FP "S" by some
quadrupolar charge operator, while others are split by a dipolar charge
operator. 
These splittings are universal, apart from some trivial potential
scattering.
This latter is generated even for small Kondo couplings 
and it becomes more pronounced as the couplings are increased.
\par

To determine the dimension of the leading irrelevant operator that governs the
new FP, we carried out a finite-size analysis of the NRG levels.
Within the NRG scheme the finite-size energy $1/L$ corresponds to
$\sim \Lambda^{-N/2}$ and the levels relax to their FP values as \cite{Delft98}
\begin{equation}
E_{\rm NRG} - E^* \propto \Lambda^{-\Delta N/2},
\end{equation}
where $E^*$ is the FP energy and $\Delta$ denotes the scaling
dimension of the leading irrelevant operator.
\begin{figure}
\begin{center}
\psfig{figure=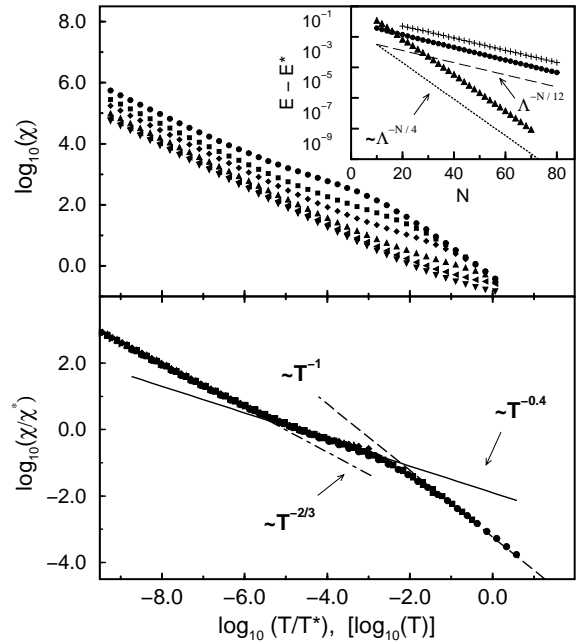,width=7.5cm}
\end{center}
\vspace*{0.05cm}
\caption{
Top: Temperature dependent magnetic susceptibility as a function of
$\log_{10} (T)$. 
\mbox{} From top to bottom, the couplings are
$4 \tilde{J}_{\rm D} = 8 \tilde{J}_{\rm Q} = 0.5$, $0.7$, $1.0$, $2.0$, $4.0$
and $8.0$.
Bottom: Rescaled magnetic susceptibility.
All the data collapse to a universal curve.
Inset: Relaxation of NRG energy levels in the vicinity of fixed points.
The top line is for $\tilde{J}_{\rm D} = \tilde{J}_{\rm Q} = 0.1$ and the
middle line is for $\tilde{J}_{\rm D} = 1.0$ and $\tilde{J}_{\rm Q} = 0$.
For comparison, we give the data for the two-channel Kondo fixed point,
scaling as $\Lambda^{-N/4}$.
We take $\Lambda = 3$ here.
}
\end{figure}
As shown in Fig.~3, for $J_{\rm Q} \neq 0$ almost the same energy level
relaxation is found as for $J_{\rm Q} = 0$
over the whole parameter space of Fig.~1, in agreement with an exponent
$\Delta =1/6$.
This implies that $\chi(T)$ and $\gamma(T)$ behave like
$T^{2 \Delta - 1} = T^{-2/3}$ at low temperatures.
Since the $S=1$ impurity spin also incorporates {\em orbital} degrees of
freedom, similar behavior is expected in the orbital (stress) susceptibility,
or the stress-induced magnetization.
\par

To determine the impurity susceptibility $\chi(T)$ we calculated the
temperature dependent magnetization induced by a small local field at the
impurity site.
The resulting curves are plotted in Fig.~3.
Each curve has an interesting region where it behaves like 
$T^{- \alpha}$ over approximately two decades, and $1/3<\alpha\approx0.4<2/3$ 
slightly depending on the magnitude of the Kondo couplings.
After adjusting the overall scale of $\chi(T)$, 
as shown in Fig.~3, we can place all the $\chi(T)$ data on a 
single universal curve using a single temperature scale 
$T^*$ (though strong coupling induces some deviations from scaling at 
higher temperatures--see the discussion below). 
This one parameter scaling in temperature strongly suggests that 
the intermediate temperature regime behavior 
reflects the new low temperature FP rather than
some unstable FP. 
For  small to intermediate 
Kondo couplings $\chi$ behaves according to the Curie-Weiss
law $T^{-\alpha}$ ($\alpha = 1$) at large temperatures.
For smaller temperatures an intermediate region appears where
$\alpha\approx 0.4$. 
As the temperature decreases further, $\chi(T)$ turns up and behaves as
$\sim T^{-2/3}$ in the vicinity of the novel FP.
For larger couplings the Curie-Weiss part is absent, and $\chi$ 
starts as $\sim T^{-1/3}$ at high temperatures and then the exponent 
$\alpha$ gradually goes up to $2/3$ at low $T$.
The appearance of the intermediate region with
$\chi(T)\sim T^{- \alpha}$ ($1/3 < \alpha < 2/3$) 
is specific to non-zero quadrupolar coupling $J_{\rm Q}$. 
When $J_{\rm Q}=0$, the dipolar coupling $J_{\rm D}$ 
gives only monotonic behavior
$T^{-1} \rightarrow T^{-2/3}$ with decreasing temperatures.
This is also a clear difference between
the Kondo effect for local triplet and doublet states.
\par

Finally, we discuss the stability of the novel FP against a
uniaxial (tetragonal) lattice distortion.
The distortion lifts the triplet degeneracy, giving either a singlet 
or doublet ground state split by a value $h_{\rm Q}$, and destabilizes
the novel triplet FP. 
Instead, we have two possible stable FP's:
when the singlet lies lowest, a Fermi Liquid FP arises, and a ground
doublet experiences a NFL FP, associated with the
two-channel quadrupolar Kondo effect for tetragonal symmetry \cite{Koga95}.
According to our NRG calculation, the crossover temperature below which the
system flows away from the cubic FP to one of the above two varies
as $h_{\rm Q}^2$ for a small distortion\cite{Koga99}.
This power implies that the operator with $j = 2$ corresponding to lattice
distortions has the same scaling dimension $1/2$ at "S" as at the unstable
FP "D" of Fig.~1.
The two tetragonal FP's are separated by a boundary line in coupling space
on which the novel cubic FP resides.
The features associated with this cubic symmetric FP are
expected to appear in some U based compounds with uniaxial anisotropy, too.
\par

In conclusion, we investigated a new Kondo Hamiltonian describing the 
dynamics of a local triplet.  The model is not restricted to uranium
ions alone, and could possibly describe praseodymium, thulium, and
terbium based materials as well.  
The quadrupolar exchange interaction drives the model to a new fixed
point, characterized by a universal charge $SU(2)$ symmetry breaking, a
leading irrelevant operator with dimension $\Delta = 1/6$, 
and a pseudo-$SU(3)$ symmetry. The universal magnetic susceptibility has an intermediate
temperature range, where it scales as $\chi \sim T^{-\alpha}$ with 
$ 1/3 < \alpha\approx 0.4 < 2/3$, with strong coupling producing a
reduction of the power law towards $\alpha=1/3$.  Since the extreme
mixed valence of UCu$_{5-x}$Pd$_x$ suggested by photoemission is
compatible with a strong coupling limit of the Kondo model, we suggest
that the intermediate temperature range susceptibility may correspond to
the ``impurity'' range identified for this material.  The surprising 
increase of the power law at lower temperatures will give the
concentrated system a greater tendency towards intersite interaction
effects, qualitatively compatible with the suggested interaction driven
low temperature physics.  To test the idea further, we strongly urge an
experimental study of this system with uranium diluted away by thorium.  
The model may prove relevant to the alloy Y$_{1-x}$U$_x$Pd$_3$, too, for
which recent neutron data suggests nearly degenerate $\Gamma_3$ and
$\Gamma_5$ states on the uranium ions \cite{Bull98}.  
\par

We thank J. W. Allen, D. E. MacLaughlin, and M. B. Maple for useful
discussions.  
This research has been supported by NSF DMR 95-28535,
the U. S - Hungarian Joint Fund Nr. 587, and grant No. DE-FG03-97ER45640 of the
U. S DOE Office of Science, Division of Materials Research.
M. K. has been supported by JSPS Postdoctoral Fellowships for Research Abroad
and G. Z. by Hungarian Grants OTKA~Nrs.~T026327 and F030041.
\par

\end{document}